\documentclass[a4paper,11pt]{article}
\usepackage{jheppub} 
\usepackage{lineno}
\usepackage{hyperref}
\usepackage{mathrsfs}  
\usepackage{latexsym,bm,amsmath,amssymb}
\usepackage{bbm}
\usepackage{physics}
\usepackage{color}
\usepackage{tikz}
\usepackage{tikz-feynman}
\usetikzlibrary{arrows}
\usepackage{graphicx,subfigure}
\usepackage{multirow}
\usepackage{cancel}
\usepackage{diagbox}
\usepackage{makecell}
 \usepackage{float}
\newcommand{\br}{{\bm r}}
\newcommand{\bk}{{\bm k}}

\newcommand{\ba}{{\bm a}}
\newcommand{\bA}{{\bm A}}

\newcommand{\bq}{{\bm q}}
\newcommand{\bx}{{\bm x}}

\newcommand{\bi}{{\mathrm{i}}}

\newcommand{\nn}{\nonumber \\}

\title{\boldmath Hall Angle of a Spatially Random Vector Model}

\author[a] {Yi-Li Wang,}
\author[a]{~Young-Kwon Han,}
\author[b]{Xian-Hui Ge,}
\author[a]{~Sang-Jin Sin}
\affiliation [a] {Department of Physics, Hanyang University,\\
	222 Wangsimni-ro, Seoul, 04763, Korea}
\affiliation[b]{ Department of Physics, Shanghai University,\\
	99 Shangda Rd., Shanghai, 200444, P.R. China}

\emailAdd{sjsin@hanyang.ac.kr}

\abstract{Strange metals exhibit linear resistivity and anomalous Hall transport, yet a comprehensive theory that accounts for both phenomena is still lacking.
	Recent studies have shown SYK-like spatially random couplings between a Fermi surface and a bosonic field, either scalar or vector type, can yield linear-$T$ resistivity. In this paper, we continue the investigation on a vector coupling in the presence of a magnetic field. We compute the fermion and boson propagators, along with the self-energy and polarization functions, and determine their dependence on the magnetic field.  Although the Hall angle does not  exhibit the signature of strange-metal, the linear-in-temperature resistivity remains at low temperatures. Results indicate that random interactions can robustly support linear transport, though additional ingredients may be required to capture the full phenomenology of strange metals.}

\begin{document}
\maketitle
\flushbottom
\section{Introduction}
Strange metal, the normal state of high-temperature superconductivity, is one of the important subject whose solution is most tantalizing  in modern physics \cite{Lee2006,Anderson2017,Lee2018,Greene2020,Varma2020,Hartnoll2022,Phillips2022}. Much effort has been made to achieve a consistent theory of strange metal, 
but there has not been a single model which produces its behavior until recently. 
A non-trivial step toward this goal has been taken recently in 
Ref.\cite{Patel2022} where  a model   giving  linear-$T$ resistivity at low temperatures  was constructed using  a spatially random coupling between a Fermi surface and a critical scalar field. The mechanism based on spatial randomness was claimed to be a `universal theory of strange metal' \cite{Patel2022}. 
The essential idea of the model \cite{Patel2022} is to consider a Yukawa coupling between electrons $\psi$ and critical scalar bosons $\phi$, $g_{ijl}(\br)\psi_i^{\dagger}(\tau,\br)\psi_j(\tau,\br)\phi_l(\tau,\br)$, such that 
\begin{eqnarray*}
	\langle g_{ijl}(\br)\rangle=0,\quad \langle g^*_{ijl}(\br)g_{i'j'l'}(\br')\rangle=g^2\delta(\br-\br')\delta_{ii',jj',ll'}.
\end{eqnarray*}
Assigning each field a flavour where $i,j,l=1,...,N$ and taking large-$N$ limit, this coupling is an analogue of the Sachdev-Ye-Kitaev (SYK) model \cite{Sachdev1993,Kitaev,Chowdhury2022}, so we can call it an `SYK-rised Yukawa model'. Such a SYK-rised scalar interaction yielded the linear resistivity at low temperatures \cite{Patel2022}. \\

Inspired by this scalar model,  we \cite{Ge:2024exw} built a vector version and found linearity  in $T$ as well. In \cite{Ge:2024exw}, a fermi surface is coupled to a \emph{vector field} $a_{\mu}$, and the interaction reads $$K_{ijl}(\br)\psi_i^{\dagger}(\tau,\br){\mathop{\nabla}\limits^{\leftrightarrow}}_{\mu}\psi_j(\tau,\br)a_l^{\mu}(\tau,\br) .$$ Strictly speaking, the scalar model and the vector model are supported by different mechanism, as the Feynman diagram for the polarisation bubble giving linear resistivity are   different. Despite this difference, 
the common origin of the starange metalicity seems to be 
the spatially random coupling between electrons and boson.\\

Although linear-$T$ resistivity, a hallmark of strange metals, has been achieved, further scrutiny is therefore required to establish the SYK-rised electron–boson coupling as a viable theory of the strange metal. According to Anderson, a theory of strange metals should also account for other anomalies \cite{Anderson1991,Anderson2013}, such as the Hall angle , \textit{inter alia}. 
Tt thus behooves us to compute the Hall conductivity.
Suppose we have a $(2+1)$-dimensional system in $x-y$ plane and a magnetic field in $z$ direction, the \emph{Hall angle} is defined as 
\begin{eqnarray}
	\tan(\Theta_H)\equiv\frac{\sigma_{xy}}{\sigma_{xx}}.
\end{eqnarray}
It has been observed that many strange metals exhibit quadratic $T$-dependence, $\cot(\Theta_H)\sim A+BT^2$ \cite{Chien1991,Tyler1997,Schrieffer2007}, first reported in 1991, where $A=0$ in pure samples. Admittedly, the quadratic Hall angle is less universal than linear resistivity. In fact, there are observations showing the breakdown of the scaling law \cite{Tyler1997,PhysRevB.53.5848} or even with the polynomial fit \cite{PhysRevB.60.R6991,Legros:2018egf,PhysRevB.95.224517}. Our major motivation in this article, however, is to test if this random coupling mechanism could correspond to some realistic materials, or equivalently, if it could reproduce other properties of at least a certain class of materials. To this end, this article continues investigating the vector model \cite{Ge:2024exw} in a magnetic field. In spite of the linear-$T$ resistivity found in \cite{Patel2022,Ge:2024exw}, the behaviour of the Hall angle does not correspond to experimental observations.\\

The paper is organised as follows.
In section 2, we offer a quick review on SYK-rised vector model and Landau basis. Section 3 illustrates how to solve the Schwinger-Dyson equations numerically. After obtaining numerical solutions, we compute the conductivity as well as Hall angle in section 4 and provide a discussion in section 5.

\section{Set-up}

\subsection{Schwinger-Dyson equation}
In this section, we give a brief introduction to spatially random vector model. For precise details (and further discussion) of this vector model, we would refer to the original paper \cite{Ge:2024exw}. \\

Let's begin with a $(2+1)$-dimensional Fermi surface coupled to a vector   field $\ba$ \cite{Ge:2024exw}, 
\begin{eqnarray}\label{eqn:action}
	S&=& \int d\tau d^2\br \left[\sum_{i=1}^N \psi^{\dagger}_i(\br,\tau)\left(\partial_\tau-\frac{\nabla^2}{2m}-\mu\right)\psi_i(\br,\tau)
	-\frac{1}{2K^2}\sum_{l=1}^N g_{ab} a_l^{a}\left(-\partial_\tau^2+\bq^2\right)a_l^{b}\right.\nonumber\\
	&&\left.
	+\sum_{i,j,l=1}^N\frac{K_{ijl}(\br)}{KN}\frac{\bi}{m}\psi^{\dagger}_i\nabla_a\psi_j\ba_l^a
	+\frac{1}{K^2N^{3/2}}\sum_{i,j,s,t=1}^N\frac{\tilde{K}_{ijst}(\br)}{2m}\ba_s\cdot\ba_t\psi_i^{\dagger} \psi_j
	\right],
\end{eqnarray}
where $\psi$ represents electrons and $a,b=\{1,2\}$ represent the spatial components. The coupling parameters $v_{ij}$, $K_{ijl}$, and $\tilde{K}_{ijst}$ obey the Gau\ss ian distribution with zero mean  and  satisfy 
\begin{eqnarray}
	&&\langle v_{ij}^*(\br)v_{i'j'}(\br')\rangle=v^2 \delta(\br-\br')\delta_{ii'}\delta_{jj'},\\
	&&\langle K_{ijl}^*(\br)K_{i'j'l'}(\br')\rangle=K^2 \delta(\br-\br')\delta_{ii'}\delta_{jj'}\delta_{ll'},\\
	&&\langle \tilde{K}_{ijst}^*(\br)\tilde{K}_{i'j's't'}(\br')\rangle=\tilde{K}^2 \delta(\br-\br')\delta_{ii'}\delta_{jj'}\delta_{ss'}\delta_{tt'}.
\end{eqnarray}
We choose $\tilde{K}=K^2$ for convenience.
\\

Suppose the fermi surface lies on the $x$-$y$ plane. Now we introduce a magnetic field $B$ along the $z$ direction and we take the Landau gauge $\bA=(-eBy,0,0)$ for   the \emph{external gauge field}. In the presence of a magnetic field, the kinetic part of electron Hamiltonian becomes $(\bk+\bA)^2/(2m)$. In other words, fermion dispersion becomes $\varepsilon_{\bk+\bA}$. Let $\bm{\pi}=\bk+\bA$ denotes the \emph{mechanical momentum}, which satisfies $[\pi_x,\pi_y]=-\bi\hbar B$.\\

In order to obtain Dyson's equations from the action \eqref{eqn:action} with $B$, one can work with the $G$-$\Sigma$ formalism \cite{Chowdhury2022}. 
The first step is to define two bi-local variables $G(x_1,x_2)$ and $D(x_1,x_2)$ as follows
\begin{eqnarray}
	&&G(x_1,x_2)\equiv-\frac{1}{N}\sum_{i}\langle\mathcal{T}\left(\psi_i(x_1)\psi_i^{\dagger}(x_2)\right)\rangle,\\
	&&D^{\mu\nu}(x_1,x_2)\equiv\frac{1}{N}\sum_l\langle\mathcal{T}\left(a_l^{\mu}(x_1)a_l^{\nu}(x_2)\right)\rangle,
\end{eqnarray}
and they will take the values of fermionic and bosonic propagators respectively at the saddle point. In addition, Lagrangian multipliers $\Sigma(x_1,x_2)$ and $\Pi(x_1,x_2)$ will be introduced to find the self-energies of electrons and bosons separately. 
By integrating out $\psi$ as well as $\phi$ and using replica trick \cite{Altland2023}, the action \eqref{eqn:action} is recast into a $G$-$\Sigma$ action \footnote{Generally, the interaction term should contain replica indices. They are omitted here, since the replica structure is unimportant in this case \cite{Chowdhury2022}. Moreover, the quenched average is equivalent to the annealed average at large-$N$ limit \cite{10.21468/SciPostPhys.14.5.113}.}
\begin{eqnarray}\label{eqn:action2}
	\frac{S}{N}&=&-\ln \det((\partial_\tau+\varepsilon_{\bk+\bA}-\mu)\delta(x-x')+\Sigma)\nonumber\\
	&&+\frac{1}{2}\ln\det(-\frac{g_{ab}}{K^2}(-\partial_\tau^2+\bq^2)\delta(x-x')-\Pi_{ab})\nonumber\\
	&&+\Tr \left(\frac{v^2}{2}G\cdot G \bar{\delta}\right)
	+\frac{1}{2m^2}\Tr\left(\frac{(\bm{\pi}_1+\bm{\pi}_2)^a(\bm{\pi}_1+\bm{\pi}_2)^b}{4}G(k_1) D_{ab}\cdot G(k_2)\bar{\delta}\right)\nonumber\\
	&&+\Tr \left(\frac{1}{8m^2}G D_{ab}D^{ab} \bar{\delta} G\right)
	-\Tr(\Sigma\cdot G)+\frac{1}{2}\Tr(\Pi^{ab}D_{ab}),
\end{eqnarray}
where $\bar{\delta}$ is the spatial delta function $\delta(\bx_1-\bx_2)$ coming from the random average, and 
\begin{equation}
	\Tr(f_1\cdot f_2)\equiv f_1^Tf_2\equiv\int dx_1dx_2f_1(x_2,x_1)f_2(x_1,x_2),
\end{equation}
with transverse $f^T(x_1,x_2)\equiv f(x_2,x_1)$ the conventions used in \cite{Guo:2022zfl,Gu:2019jub}.
At the saddle point, from 
\begin{eqnarray}
	0&=&\frac{\delta S}{N}\nn
	&\equiv&
	\Tr(\delta\Sigma(G_*[\Sigma]-G)+\delta G(\Sigma_*[G]-\Sigma) 
	+\frac{1}{2}\delta\Pi_{ab}(D^{ab}-{D_*}^{ab}[\Pi^{ab}])+\delta D_{ab}(\Pi^{ab}-\Pi_*^{ab}[D_{11}])),\nn
\end{eqnarray}
 one obtains the Dyson's equations
\begin{align}
	&G=G_*=\left(-\partial_\tau-\varepsilon_{\bk+\bA}+\mu-\Sigma\right)^{-1},\label{eqn:f-propagator}\\
	&\Sigma=\Sigma_*=
	\frac{(\bm{\pi}_1+\bm{\pi}_2)^a(\bm{\pi}_1+\bm{\pi}_2)^b}{4m^2}D_{ab} G\bar{\delta}+\frac{1}{4m^2}D_{ab}D^{ab}G\bar{\delta},\label{eqn:f-selfenergy}\\
	&D_{ab}={D_*}_{ab}=K^2(-g^{ab}(-\partial_\tau^2+\bq^2)-K^2\Pi^{ab})^{-1},\label{eqn:b-propagator}\\
	&\Pi_{ab}={\Pi_*}_{ab}=-\frac{(\bm{\pi}_1+\bm{\pi}_2)_a(\bm{\pi}_1+\bm{\pi}_2)_b}{4m^2}G\bar{\delta}\cdot G
	-\frac{1}{4m^2}G D_{ab}G\bar{\delta}.\label{eqn:b-selfenergy}
\end{align}
We aim to solve self-consistent equations \eqref{eqn:f-propagator}-\eqref{eqn:b-selfenergy}.

\subsection{Landau-level basis}
In a magnetic field, there will be an orbit quantization and it is more convenient to work in Landau-level basis. This section only contains a summary of the conventions used in this article and we refer readers interested in quantum field theories in a magnetic field to an elaborate note \cite{Miransky2015} by Miransky and Shovkovy.\\

The building blocks of Landau levels are ladder operators defined as
\begin{eqnarray}
	\mathfrak{a}=\frac{1}{2e\hbar B}(\pi_x-\bi\pi_y),\quad \mathfrak{a}^\dagger=\frac{1}{2e\hbar B}(\pi_x+\bi\pi_y),
\end{eqnarray}
satisfying $[\mathfrak{a},\mathfrak{a}^\dagger]=1$. The Hamiltonian thus becomes
\begin{eqnarray}
	H=\hbar\omega_B\left(\mathfrak{a}^\dagger\mathfrak{a}+\frac{1}{2}\right),
\end{eqnarray}
where $\omega_B=eB/m$ is the \emph{cyclotron frequency}. The Hilbert space is built with the ground state $\ket{0}$ obeying $\mathfrak{a}\ket{0}=0$, and the rest states satisfying
\begin{equation}
	\mathfrak{a}^\dagger\ket{n}=\sqrt{n+1}\ket{n+1},\quad \mathfrak{a}\ket{n}=\sqrt{n}\ket{n-1}.
\end{equation}
A state $\ket{n}$ is characterised by its eigen-energy
\begin{eqnarray}
	E_n=\hbar\omega_B\left(n+\frac{1}{2}\right).
\end{eqnarray}
The energy levels are named \emph{Landau levels}. A state $\ket{n}$ has degeneracy $g_L=eBS/(2\pi\hbar)$, with $S$ the area of the sample.\\

Since the gauge field breaks translational symmetry in $y$ direction, the electron wavefunction is a plane wave only in $x$ direction with momentum $k$. 
The full electron wavefunction $\psi_{nk}$ (in coordinate basis) reads \cite{Miransky2015}
\begin{eqnarray}
	\psi_{nk}(x,y)\equiv\bra{\bx}\ket{n}=\frac{1}{\sqrt{2\pi\ell_B}}\frac{e^{-(y-k\ell_B)^2/2\ell_B^2}}{\sqrt{2^n n!\sqrt{\pi}}}H_n\left(\frac{y}{\ell_B}-k\ell_B\right)e^{\bi kx},
\end{eqnarray}
where $H_n(z)$ is the Hermite polynomial function. The \emph{magnetic length} $\ell_B=\sqrt{\hbar/(eB)}$ characterise the scale governing quantum phenomena in a magnetic field. The normalisation and completeness conditions are
\begin{eqnarray}
	&&\int d\bx\bra{n}\ket{\bx}\bra{\bx}\ket{n'}=\int dx dy \psi_{nk}^*(x,y)\psi_{n'k'}(x,y)=\delta_{nn'}\delta(k-k'),\\
	&&\sum_{n=0}^{\infty}\int_{-\infty}^{+\infty}dk \bra{u}\ket{n}\bra{n}\ket{u}=
	\sum_{n=0}^{\infty}\int_{-\infty}^{+\infty}dk\psi_{nk}(\bx)\psi_{nk}^*(\bx')=\delta(\bx-\bx').
\end{eqnarray}
Here $\ket{n}=\ket{n;k}$, but we drop $k$ for convenience. \\

It is helpful to introduce the \emph{velocity operator} $\bm{v}\equiv\bm{\pi}/m$ as follows \cite{Guo2024}
\begin{eqnarray}
	&&\bm{v}_x=\frac{(\bk_x-eBy)}{m}=-\frac{\mathfrak{a}+\mathfrak{a}^{\dagger}}{\sqrt{2}m\ell_B},\\
	&&\bm{v}_y=\frac{\bk_y}{m}=\frac{\bi(-\mathfrak{a}+\mathfrak{a}^{\dagger})}{\sqrt{2}m\ell_B}
\end{eqnarray}
Their matrix elements then reads
\begin{eqnarray}
	&&\mathscr{V}^x_{n'k',nk}=\bra{n',k'}\bm{v}_x\ket{n,k}=-\frac{1}{\sqrt{2}m\ell_B}\left(\sqrt{n}\delta_{n,n'+1}+\sqrt{n'}\delta_{n',n+1}\right)2\pi\delta(k-k'),\\
	&&\mathscr{V}^y_{n'k',nk}=\bra{n',k'}\bm{v}_y\ket{n,k}=\frac{\bi}{\sqrt{2}m\ell_B}\left(\sqrt{n'}\delta_{n',n+1}-\sqrt{n}\delta_{n,n'+1}\right)2\pi\delta(k-k').
\end{eqnarray}
Having reviewed the Landau-level basis, let us continue to solve the Schwinger-Dyson's equation.

\section{Dyson's Equations}
\subsection{Propagators and self-energies in Landau level basis}\label{sec:dyson}

\paragraph{Electron propagator}
It is more convenient to solve Dyson's equations in Landau-level basis. Let us begin with the electron propagator, the simplest one among all quantities to be found. In coordinate space, the fermionic propagator reads 
\begin{eqnarray}
	G(x,x')&=&(-\partial_{\tau}+\mu-\varepsilon_{\bk+\bA}-\Sigma)^{-1}(x,x')\nn
	&=&\bra{\bx}(\bi\omega+\mu-\hat{H}-\Sigma)^{-1}\ket{\bx'}
\end{eqnarray}
in terms of Matsubara frequencies $\omega$, where $x=(t,\bx)$. This can be decomposed into Landau-level basis
\begin{eqnarray}
	G(\bx,\bx';\bi\omega)&=&\bra{\bx}(\bi\omega+\mu-\hat{H}-\Sigma)^{-1}\ket{\bx'}\nn
	&=&\sum_{n=0}^{\infty}\int_{-\infty}^{+\infty}\frac{dk}{2\pi}
	\bra{\bx}(\bi\omega+\mu-\hat{H}-\Sigma)^{-1}\ket{nk}\bra{nk}\ket{\bx'}\nn
	&=&\sum_{n=0}^{\infty}\int_{-\infty}^{+\infty}\frac{dk}{2\pi}
	\bra{\bx}\ket{nk}(\bi\omega+\mu-\omega_B(n+\frac{1}{2})-\Sigma)^{-1}\bra{nk}\ket{\bx'}\nn
	&\equiv&\sum_{n=0}^{\infty}\int_{-\infty}^{+\infty}\frac{dk}{2\pi}
	\psi_{nk}(\bx)\psi_{nk}(\bx')G_n(\bi\omega),
\end{eqnarray}
where
\begin{eqnarray}\label{eqn:npropa2}
	G_n(\bi\omega)=\frac{1}{\bi\omega+\mu-\omega_B(n+\frac{1}{2})-\Sigma}.
\end{eqnarray}
In \eqref{eqn:npropa2}, we assume that the self-energy $\Sigma=\Sigma(\bi\omega)$ does not depend on momentum and it is diagonal in Landau-level basis. The interaction $\psi^{\dagger}\nabla_a\psi\ba^a$ encodes information of fermionic momenta. Hence, the self-enegies should depend on momentum, even with a spatial contraction, as is shown by eqn.\eqref{eqn:f-selfenergy}. To render the computation manageable, we assume that only particles near Fermi surface contribute. This means that we will take the \emph{typical Landau level} on each vertex. The typical value of Landau level $n_t$ satisfies $n\omega_B\simeq k_F^2/2m$, which gives \cite{Guo2024} 
\begin{eqnarray}
	n_t\simeq (k_F\ell_B)^2/2.
\end{eqnarray}
Though the critical value of $n_t$, beyond which the substitution is justifiable, cannot be reliably estimated at this stage, the least requirement that must be met is $\omega_B<k_F^2/2m$.
Later we will show that the self-energy $\Sigma$ is diagonal in Landau-level basis. Using the fact that \cite{Gradshtein2015}
\begin{eqnarray}
	\int_{-\infty}^{\infty}dx e^{-x^2}H_m(x+y)H_n(x+z)dx=2^n\pi^{1/2}m!z^{n-m}L_{m}^{n-m}(-2yz),
\end{eqnarray}
where $L_m^n$ is the Laguerre polynomial, and letting
\begin{eqnarray}
	k\ell_B-\frac{\bi(y-x)+(y'+x')}{2\ell_B}=\tilde{k}\ell_B,
\end{eqnarray}
one finds
\begin{eqnarray}
	G(\bx,\bx';\bi\omega)=\sum_nG_n(\bi\omega)\frac{e^{\bi\theta_B(\bx,\bx')}}{2\pi\ell_B^2}
	\exp(-\frac{|\bx-\bx'|^2}{4\ell_B^2})L_n\left(\frac{|\bx-\bx'|^2}{2\ell_B^2}\right),
\end{eqnarray}
with $\theta_B(\bx,\bx')=(y-x)(y'+x')/2\ell_B^2$. We thus obtain the electron propagator in a magnetic field.\\

\paragraph{Boson self-energy}
Having known the electron propagator, it is straightforward to find the bosonic self-energy $\Pi(\bi\Omega)$ according to \eqref{eqn:b-selfenergy}. Due to spatial randomness, the boson self-energy has no dependence on canonical momentum either. Here we only consider the diagrams up to one loop, say $G\cdot G$ graph according to eqn.\eqref{eqn:b-selfenergy}, and we find
\begin{eqnarray}\label{eqn:pigeneral}
	-\Pi^{\mu\nu}(\bi\Omega) &=&\sum_{\omega}\int d\bx d\bx'dk dk' \sum_{n,n'}e^{-\bi\bq(\bx-\bx')}\delta(\bx-\bx')\nn
	&&\times\bra{n';k'}\ket{\bx}\bra{\bx}\bm{v}^{\mu}\hat{G}(\bi\omega)\ket{n;k}\bra{n;k}\ket{\bx'}\bm{v}^{\nu}
	\bra{\bx'}\hat{G}(\bi(\omega-\Omega))\ket{n';k'}.\nn
\end{eqnarray}
Now let us first find $\Pi^{xx}$, 
\begin{eqnarray}
	-\Pi^{xx}(\bi\Omega) &=&\sum_{\omega}\int d\bx d\bx'dk dk' \sum_{n,n'}e^{-\bi\bq(\bx-\bx')}\delta(\bx-\bx')\nn
	&&\times\bra{n';k'}\ket{\bx}\bra{\bx}\bm{v}^{x}\hat{G}(\bi\omega)\ket{n;k}\bra{n;k}\ket{\bx'}\bm{v}^{x}
	\bra{\bx'}\hat{G}(\bi(\omega-\Omega))\ket{n';k'}\nn
	&=&\frac{1}{2m^2\ell_B^2}\sum_{\omega}\int d\bx dk dk' \sum_{n,n'} G_n(\bi\omega) G_{n'}(\bi(\omega-\Omega))\nn
	&&\left(\sqrt{n'+1}\psi^*_{n'+1,k'}(\bx)\psi_{n,k}(\bx)+\sqrt{n+1}\psi^*_{n',k'}(\bx)\psi_{n+1,k}(\bx)\right)\nn
	&&\left(\sqrt{n+1}\psi^*_{n+1,k}(\bx)\psi_{n',k'}(\bx)+\sqrt{n'+1}\psi^*_{n,k}(\bx)\psi_{n'+1,k'}(\bx)\right)
\end{eqnarray}
Using the formula \cite{Gradshtein2015}
\begin{eqnarray}
	\int e^{-x^2}H_n(x)H_m(x)dx=\delta_{m,n} 2^n n!\sqrt{\pi},
\end{eqnarray}
we find 
\begin{eqnarray}\label{eqn:pixx}
		&&-\Pi^{xx}(\bi\Omega)=-\Pi^{yy}(\bi\Omega) \nn
		&=&\frac{S}{2m^2\ell_B^2}\sum_{\omega}\sum_{n,n'}\left(\frac{n'+1}{4\pi^2 \ell_B^4}G_n(\bi\omega) G_{n'}(\bi(\omega-\Omega))+\frac{n+1}{4\pi^2 \ell_B^4}G_n(\bi\omega) G_{n'}(\bi(\omega-\Omega))\right)\nn
		&\simeq&\frac{S}{2m^2\ell_B^2}\frac{k_F^2}{4\pi^2\ell_B^2}
		\sum_{\omega}\sum_{n,n'}G_n(\bi\omega) G_{n'}(\bi(\omega-\Omega),
\end{eqnarray}
where we have substitute the typical Landau level $n_t$ into the summation.
Similarly,
\begin{eqnarray}\label{eqn:pixy}
		&&-\Pi^{xy}(\bi\Omega)=\Pi^{yx}(\bi\Omega) \nn
		&=&\frac{\bi S}{2m^2\ell_B^2}\sum_{\omega}\sum_{n,n'}\left(-\frac{n}{4\pi^2 \ell_B^4}G_n(\bi\omega) G_{n'}(\bi(\omega-\Omega))+\frac{n+1}{4\pi^2 \ell_B^4}G_n(\bi\omega) G_{n'}(\bi(\omega-\Omega))\right)\nn
		&=&\frac{\bi S}{2m^2\ell_B^2}\sum_{\omega}\sum_{n,n'}\frac{1}{4\pi^2 \ell_B^4}G_n(\bi\omega) G_{n'}(\bi(\omega-\Omega)).
\end{eqnarray}
Here we find that the off-diagonal components receive an extra factor of $1/\ell_B^2$, compared with the longitudinal parts. The interaction $(\psi^{\dagger}\nabla_{\mu}\psi) \bm{a}^{\mu}$ introduces $\bm{v}^{\mu}\bm{v}^{\nu}$ to the vertices of self-energies. Eqn.\eqref{eqn:pixy} shows that $\bm{v}^{x}$ and $\bm{v}^{y}$ partly cancel contributions from each other. On the other hand, the diagonal component \eqref{eqn:pixx} does not observe such cancellation. Instead, we take $n=n_t\simeq (k_F\ell_B)^2/2$ on each vertex in the last step of eqn.\eqref{eqn:pixx}. Consequently, one obtains $|\Pi_{xy}|\ll\Pi_{xx}$.

\paragraph{Boson propagator}
Knowing boson self-energy, one immediately finds boson propagator. According to eqn.\eqref{eqn:b-propagator}, the boson propagator writes
\begin{eqnarray}
	D_{ab}=
	K^2\begin{pmatrix}
		(\Omega^2+\bq^2)-K^2\Pi^{xx}(\bi\Omega) & -K^2\Pi^{xy}(\bi\Omega) \\
		K^2\Pi^{xy}(\bi\Omega) & (\Omega^2+\bq^2)-K^2\Pi^{xx}(\bi\Omega) 
	\end{pmatrix}^{-1},
\end{eqnarray}
and one finds
\begin{align}
	D_{xx}=D_{yy}&=K^2\frac{(\Omega^2+\bq^2)-K^2\Pi^{xx}(\bi\Omega)}{(\Omega^2+\bq^2-K^2\Pi^{xx}(\bi\Omega))^2+K^4\Pi^{xy}(\bi\Omega)^2}\nn&\simeq K^2\frac{1}{\Omega^2+\bq^2-K^2\Pi^{xx}(\bi\Omega)},\label{eqn:dxx}\\
    D_{xy}=-D_{yx}&=K^2\frac{K^2\Pi^{xy}(\bi\Omega)}{(\Omega^2+\bq^2-K^2\Pi^{xx}(\bi\Omega))^2+K^4\Pi^{xy}(\bi\Omega)^2}\nn
    &\simeq K^2\frac{K^2\Pi^{xy}(\bi\Omega)}{(\Omega^2+\bq^2-K^2\Pi^{xx}(\bi\Omega))^2}.\label{eqn:dxy}
\end{align}
The approximations in the last steps above are made as $|\Pi_{xy}|\ll |\Pi_{xx}|$. 
The $xx-$ and $xy-$components only differ by their numerators. The one of $xx$ component receives contributions from the bosonic kinetic kernel, while that of the $xy$ component only contains self-energy $\Pi_{xy}$. In addition, we already  argued that $|\Pi_{xy}|\ll|\Pi_{xx}|$.
In total, the longitudinal components of all self-energies and propagators are larger than the off-diagonal components. That is, $|D_{xy}|\ll|D_{xx}|$.

\paragraph{Electron self-energy}
For vector coupling, eqn.\eqref{eqn:f-selfenergy} describes the electron self-energy, where $\Sigma(\bx,\bx')=\bm{v}^a\bm{v}^bG(\bx,\bx')D_{ab}(\bx,\bx')\equiv\Sigma^{xx}+\Sigma^{xy}$.
We may first assume that $\Sigma$ is not diagonalised in Landau-level basis. The $xx$ component then reads
\begin{eqnarray}\label{eqn:sigmaxx}
	&&\Sigma^{xx}_{n''k'',n'k'}(\bx,\bx')\nn
	&=&2T\sum_{\Omega}\int  d\bx d\bx' \frac{dk}{2\pi}\frac{d^2\bq}{4\pi^2}\sum_n G_n(\bi(\omega+\Omega))D_{xx}(\bq,\bi\Omega)e^{-\bi(\bx'-\bx)}\delta(\bx-\bx')\nn
	&&\bra{n'';k''}\ket{\bx}\bra{\bx}\bm{v}^x\ket{n}\bra{n}\ket{\bx'}\bm{v}^x\bra{\bx'}\ket{n';k'}\nn
	&=&\frac{1}{m^2\ell_B^2}\sum_{\Omega}\int d\bx \frac{dk}{2\pi}\frac{d^2\bq}{4\pi^2}T\sum_n
	G_n(\bi(\omega+\Omega))D_{xx}(\bq,\bi\Omega)\nn
	&&(n\phi^*_{n'';k''}(\bx)\phi_{n-1;k}(\bx)\phi^*_{n-1;k}(\bx)\phi_{n';k'}(\bx)
	+(n+1)\phi^*_{n'';k''}(\bx)\phi_{n+1;k}(\bx)\phi^*_{n+1;k}(\bx)\phi_{n';k'}(\bx))\nn
	&=&\frac{1}{m^2\ell_B^2}\frac{1}{2\pi\ell_B^2}T\sum_{\Omega}\int\frac{d^2\bq}{4\pi^2}\sum_n
	G_n(\bi(\omega+\Omega))D_{xx}(\bq,\bi\Omega)(2n+1)\delta_{n'',n'}\delta(k''-k')\nn
	&\simeq&\frac{k_F^2}{2\pi m^2\ell_B^2}T\sum_{\Omega}\int\frac{d^2\bq}{4\pi^2}\sum_n
	G_n(\bi(\omega+\Omega))D_{xx}(\bq,\bi\Omega)\delta_{n'',n'}\delta(k''-k').
\end{eqnarray}
For $xy$ components, one obtains
\begin{eqnarray}\label{eqn:sigmaxy}
	&&\Sigma^{xy}_{n''k'',n'k'}(\bx,\bx')\nn
	&=&2T\sum_{\Omega}\int  d\bx d\bx' \frac{dk}{2\pi}\frac{d^2\bq}{4\pi^2}\sum_n G_n(\bi(\omega+\Omega))D_{xy}(\bq,\bi\Omega)e^{-\bi(\bx'-\bx)}\delta(\bx-\bx')\nn
	&&\bra{n'';k''}\ket{\bx}\bra{\bx}\bm{v}^x\ket{n}\bra{n}\ket{\bx'}\bm{v}^y\bra{\bx'}\ket{n';k'}\nn
	&=&\frac{\bi}{m^2\ell_B^2}T\sum_{\Omega}\int d\bx \frac{dk}{2\pi}\frac{d^2\bq}{4\pi^2}\sum_n
	G_n(\bi(\omega+\Omega))D_{xy}(\bq,\bi\Omega)\nn
	&&(-n\phi^*_{n'';k''}(\bx)\phi_{n-1;k}(\bx)\phi^*_{n-1;k}(\bx)\phi_{n';k'}(\bx)
	+(n+1)\phi^*_{n'';k''}(\bx)\phi_{n+1;k}(\bx)\phi^*_{n+1;k}(\bx)\phi_{n';k'}(\bx))\nn
	&=&\frac{\bi}{m^2\ell_B^2}\frac{1}{2\pi\ell_B^2}T\sum_{\Omega}\int\frac{d^2\bq}{4\pi^2}\sum_n
	G_n(\bi(\omega+\Omega))D_{xy}(\bq,\bi\Omega)\delta_{n'',n'}\delta(k''-k').
\end{eqnarray}
Hence both $\Sigma_{xx}$ and $\Sigma_{xy}$ is diagonal in Landau-level basis and it is proportional to the identity matrix. We will thus drop the dependence on $(n,k)$ indices. Moreover, one finds $|\Sigma_{xy}|\ll|\Sigma_{xx}|$ for the same reason analysed when we compute boson self-energies $\Pi^{\mu\nu}$, so we will ignore $\Sigma_{xy}$ in the following computation.

\subsection{Numerical solutions}
\subsubsection{Auxiliary propagators}
The equations above are too cumbersome to be solved analytically, so we will solve them numerically using the strategy introduced in \cite{Schmalian1996}. As is shown in section \ref{sec:dyson}, the spatial delta make all momentum to be integrated out individually. To simplify the computation, we can define auxiliary Green's functions $\bar{G}(\bi\omega)$ and $\bar{D}_{\mu\nu}(\bi\Omega)$. Firstly, the auxiliary electron propagator $\bar{G}(\bi\omega)$ is defined as
\begin{align}
	&\bar{G}(\bi\omega)\equiv\sum_nG_n(\bi\omega).
\end{align}
In order to perform a numerical calculation, we put a cutoff on Landau levels $n$, such that $n_-\leq n \leq n_+-1$. Following the strategy in \cite{Guo2024}, we choose $-n_-=n_+=W/(2\omega_B)$, where $W$ is interpreted as the bandwidth. Therefore,
\begin{eqnarray}\label{eqn:gbar}
	\bar{G}(\bi\omega)&=&\sum_{n=-n_+}^{n_+-1}\frac{1}{\bi\omega+\mu-\omega_B(n+\frac{1}{2})-\Sigma(\bi\omega)}\nn
	&=&\frac{1}{\omega_B}\left[\psi\left(\frac{1}{2}-n_+-\frac{\bi\omega+\mu-\Sigma(\bi\omega)}{\omega_B}\right)
	-\psi\left(\frac{1}{2}+n_+-\frac{\bi\omega+\mu-\Sigma(\bi\omega)}{\omega_B}\right)\right].
\end{eqnarray}
Using \eqref{eqn:dxx} and \eqref{eqn:dxy}, the auxiliary Boson Green's functions $\bar{D}$ are
\begin{align}
	\bar{D}_{xx}(\bi\Omega)&\equiv K^2\int \frac{d^2\bq}{(2\pi)^2}\frac{(\Omega^2+\bq^2)-K^2\Pi^{xx}(\bi\Omega)}{(\Omega^2+\bq^2-K^2\Pi^{xx}(\bi\Omega))^2+K^4\Pi^{xy}(\bi\Omega)^2}\nn
	&\simeq \frac{K^2}{4\pi}\ln(\frac{\Omega^2+\Lambda_q^2-K^2\Pi^{xx}(\bi\Omega)}{\Omega^2-K^2\Pi^{xx}(\bi\Omega)}),\label{eqn:dbarxx}\\
	\bar{D}_{xy}(\bi\Omega)&\equiv K^2\int\frac{d^2\bq}{(2\pi)^2} \frac{K^2\Pi^{12}(\bi\Omega)}{(\Omega^2+\bq^2-K^2\Pi^{11}(\bi\Omega))^2+K^4\Pi^{12}(\bi\Omega)^2}\nn
	&\simeq K^2\int\frac{d^2\bq}{(2\pi)^2} \frac{K^2\Pi^{12}(\bi\Omega)}{(\Omega^2+\bq^2-K^2\Pi^{11}(\bi\Omega))^2}\nn
	&=\frac{K^2}{4\pi}\frac{K^2\Pi^{xy}(\bi\Omega)}{\Omega^2-K^2\Pi^{xx}(\bi\Omega)}.\label{eqn:dbarxy}
\end{align}
where $\Lambda_q$ is a bosonic momentum cutoff. During the numerical calculation, we will neglect $\bar{D}_{xy}$ since it is much smaller than $\bar{D}_{xx}$. 
After obtaining the solutions, we can input the data to find $\Pi_{xy}$ \eqref{eqn:pixy} and $D_{xy}$ \eqref{eqn:dxy} (thus $\bar{D}_{xy}$ \eqref{eqn:dbarxy} as well). 
Moreover, we will subtract the thermal fluctuations, so $\Pi_{ab}(\Omega)$ will be replaced by $\Pi_{ab}(\Omega)-\Pi_{ab}(0)$.\\

\subsubsection{The solution}\label{subsec:solution}
We follow the strategy introduced in \cite{Schmalian1996} to perform a numerical calculation.\footnote{Full details can be found in Appendix.\ref{app:summation}.} 
\begin{itemize}
	\item[1.] We begin with the auxiliary electron propagator and do an analytical continuation $\bi\omega\to\omega+\bi\eta$ ($\eta\in\mathbb{R}^+$ and $\eta\ll 1$), so $\bar{G}$ reads
	\begin{eqnarray}\label{eqn:gnum}
		&&\bar{G}(\omega)=\frac{1}{\omega_B}\left[\psi\left(\frac{1}{2}-n_+-\frac{\omega+\bi\eta+\mu-\Sigma(\omega)}{\omega_B}\right)
		-\psi\left(\frac{1}{2}+n_+-\frac{\omega+\bi\eta+\mu-\Sigma(\omega)}{\omega_B}\right)\right].\nn
	\end{eqnarray}
	 We will arbitrarily choose the initial value of the electron propagator, $\bar{G}_{\text{i}}$.
	\item[2.] We Fourier transform ($F(t)=\int d\omega f(\omega)\exp(-\bi\omega t)/2\pi$) this  initial value $\bar{G}_i(\omega)$ to $\bar{G}_i(t)$ and compute the boson self-energies
	\begin{align}
		&\Pi_{xx}(t)=\Pi_{yy}(t)=-\frac{S}{m^2\ell_B^2}\frac{k_F^2}{4\pi^2\ell_B^2}\Re{\tilde{\bar{G}}^*(t)\bar{G}(t)},
	\end{align}
	where $\tilde{\bar{G}}(\omega)\equiv-2n_F(\omega)\Im{\bar{G}(\omega)}$.
	Transforming the result back to momentum representation and getting $\Pi_{xx}(\Omega)$, one obtains the auxiliary propagator
	\begin{eqnarray}
		\bar{D}_{xx}(\Omega)
		&=&\frac{K^2}{4\pi}\ln(\frac{(\eta-\bi\Omega)^2+\Lambda_q^2-K^2\Pi^{xx}(\Omega)}{(\eta-\bi\Omega)^2-K^2\Pi^{xx}(\Omega)}).
	\end{eqnarray}
	\item[3.] Next one transforms $\bar{D}_{xx}(\Omega)$ to $\bar{D}_{xx}(t)$, and the electron self-energy reads
	\begin{eqnarray}
		&&\Sigma(t)\simeq-\frac{k_F^2}{\pi m^2\ell_B^2}\left(\tilde{\bar{D}}_{xx}^*(t)\bar{G}(t)+\tilde{\bar{G}}(t)\bar{D}_{xx}(-t)
		\right)
	\end{eqnarray}
	where $\tilde{\bar{D}}_{xx}(\omega)=-n_B(\omega)\Im{\bar{D}_{xx}(\omega)}$. Fourier transforming $\Sigma(t)$ yields $\Sigma(\omega)$.
	\item[4.] Using this $\Sigma(\omega)$ to obtain a new value of $\bar{G}_{\text{new}}$ according to \eqref{eqn:gnum}. If $\bar{G}_{\text{new}}$ converges to $\bar{G}_{\text{i}}$, stop. Otherwise, set $\bar{G}_{\text{new}}$ as the new $\bar{G}_{\text{i}}$ and repeat step 2-4 until convergence happens. One thus gets the  numerical solution.
	\item[5.] Applying the solutions of $G$ to obtain
	\begin{align}
		&\Pi_{xy}(t)=-\Pi_{yx}(t)
		=-\frac{\bi S}{(m^2\ell_B^2)(4\pi^2\ell_B^4)}
		\Re{\tilde{\bar{G}}^*(t)\bar{G}(t)},
	\end{align}
	and thus
	\begin{eqnarray}
		\bar{D}_{xy}(\Omega)&=&\frac{K^2}{4\pi}\frac{K^2\Pi^{xy}(\Omega)}{(\eta-\bi\Omega)^2-K^2\Pi^{xx}(\Omega)}.
	\end{eqnarray}
	One hitherto obtains the complete numerical solutions to Dyson's equations.
\end{itemize}

\begin{figure}[H]
	\centering
	\includegraphics[width=0.6\textwidth]{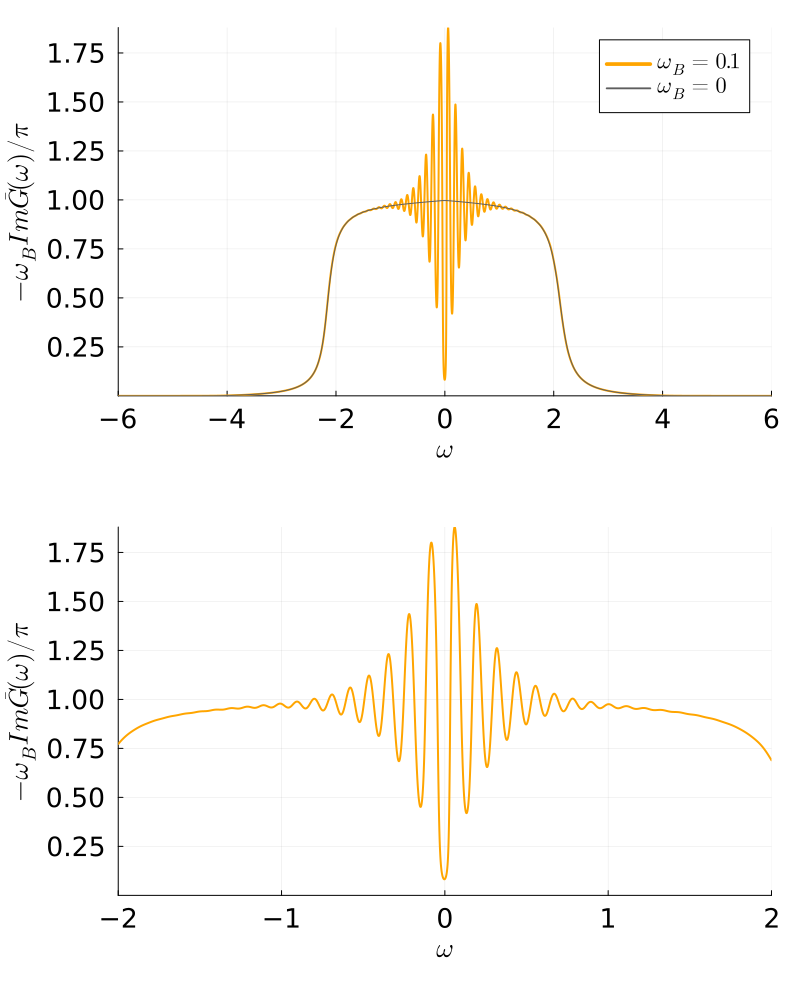}
	\caption{Density of state of electrons. The peaks correspond to Landau levels. \label{fig:gpropagator}}
\end{figure}

\begin{figure}[H]
	\centering
	\includegraphics[width=0.6\textwidth]{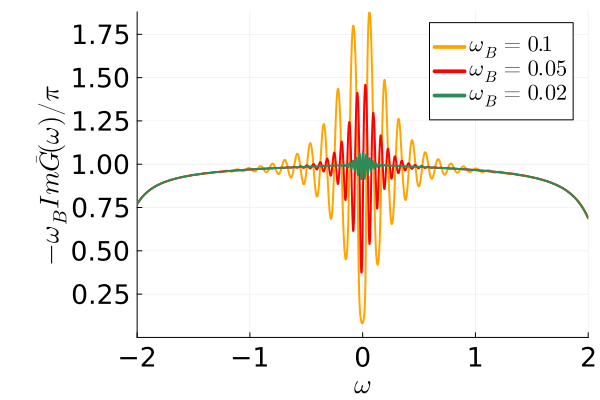}
	\caption{Numerical solutions with different cyclotron frequencies $\omega_B$. The amplitude of oscillation becomes smaller as magnetic field declines. The orange, red, and green lines stands for solutions with $\omega_B=0.1,0.05,0.02$ respectively. \label{fig:omegab}}
\end{figure}

\begin{figure}[H]
	\centering
	\includegraphics[width=0.5\textwidth]{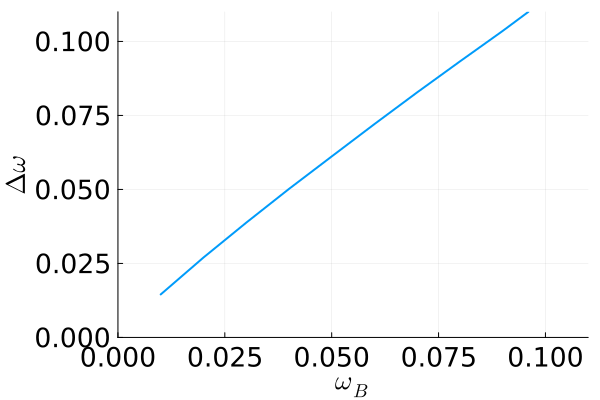}
	\caption{The relation between average separation of minimum and the cyclotron frequency $\omega_B$.\label{fig:level2}}
\end{figure}

\begin{figure}[H]
	\centering
	\includegraphics[width=0.5\textwidth]{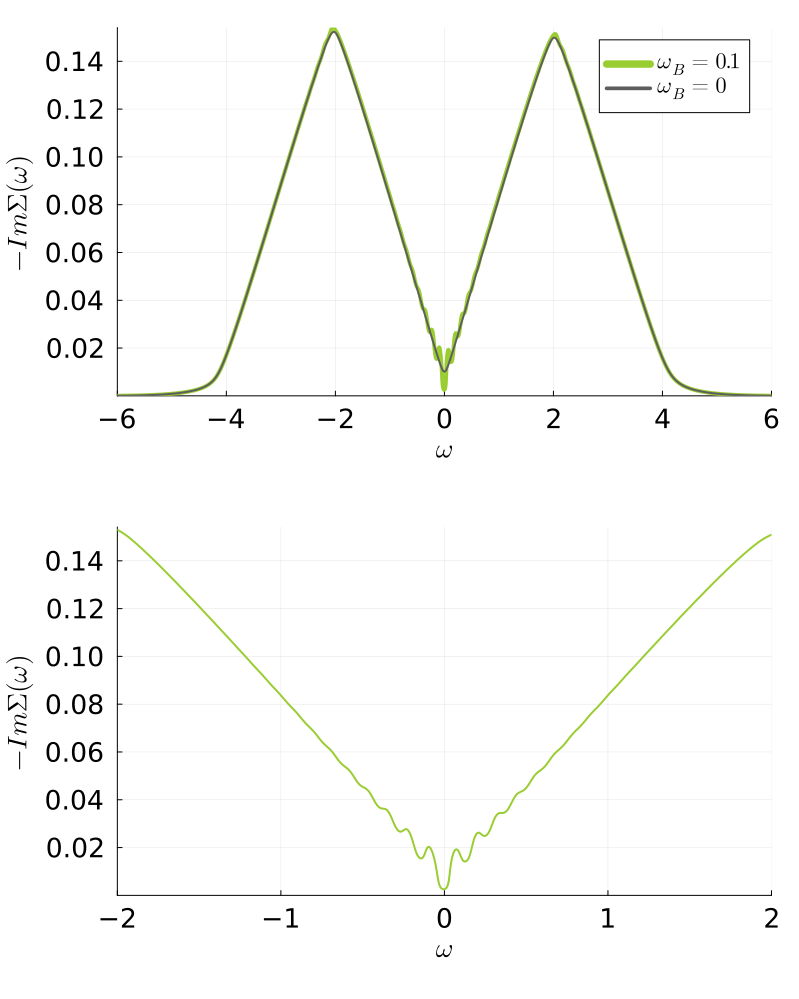}
	\caption{The Green thick line illustrates the electron self-energy $\Sigma(\omega)$ when $\omega_B=0.1$. It takes a similar form with the one where $\omega_B=0$, which is represented by the grey line. The approximate linearity (except small oscillations) implies a linear resistivity at low temperatures.  \label{fig:sigmaenergy}}
\end{figure}

\begin{figure}[H]
	\centering
	\includegraphics[width=0.47\textwidth]{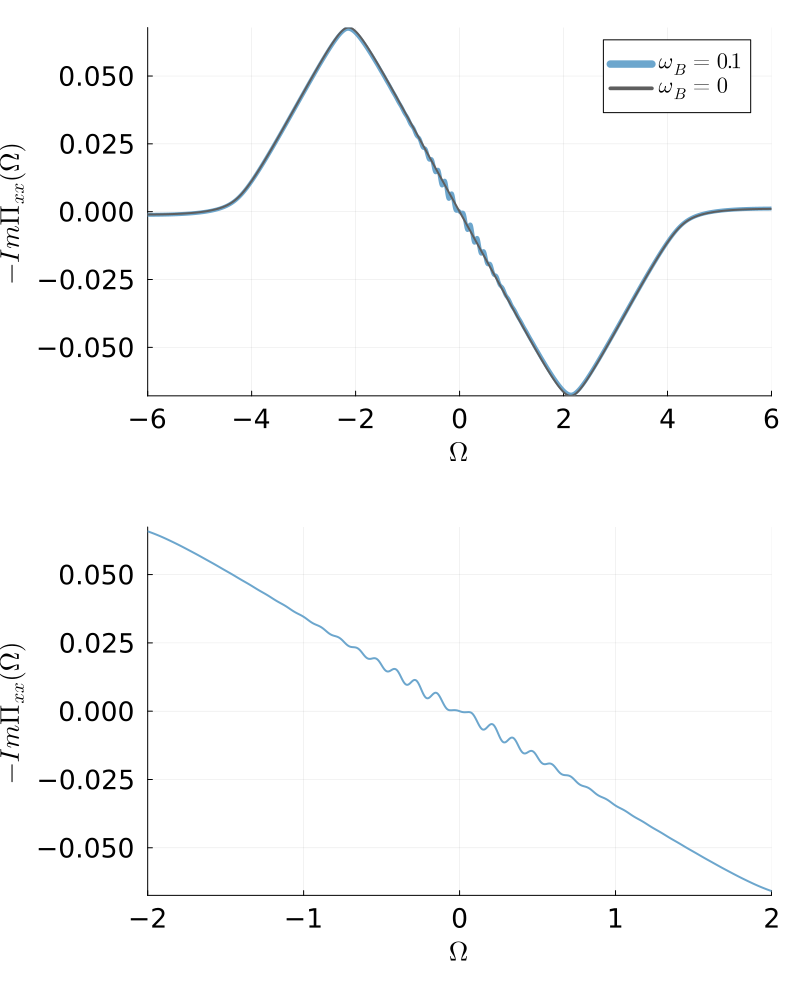}
	\caption{Numerical solutions of $\Pi_{xx}(\Omega)$. The blue line represents $\omega_B=0.1$, and the grey line shows zero-field solution.\label{fig:pixx}}
\end{figure}

\begin{figure}[H]
	\centering
	\includegraphics[width=0.45\textwidth]{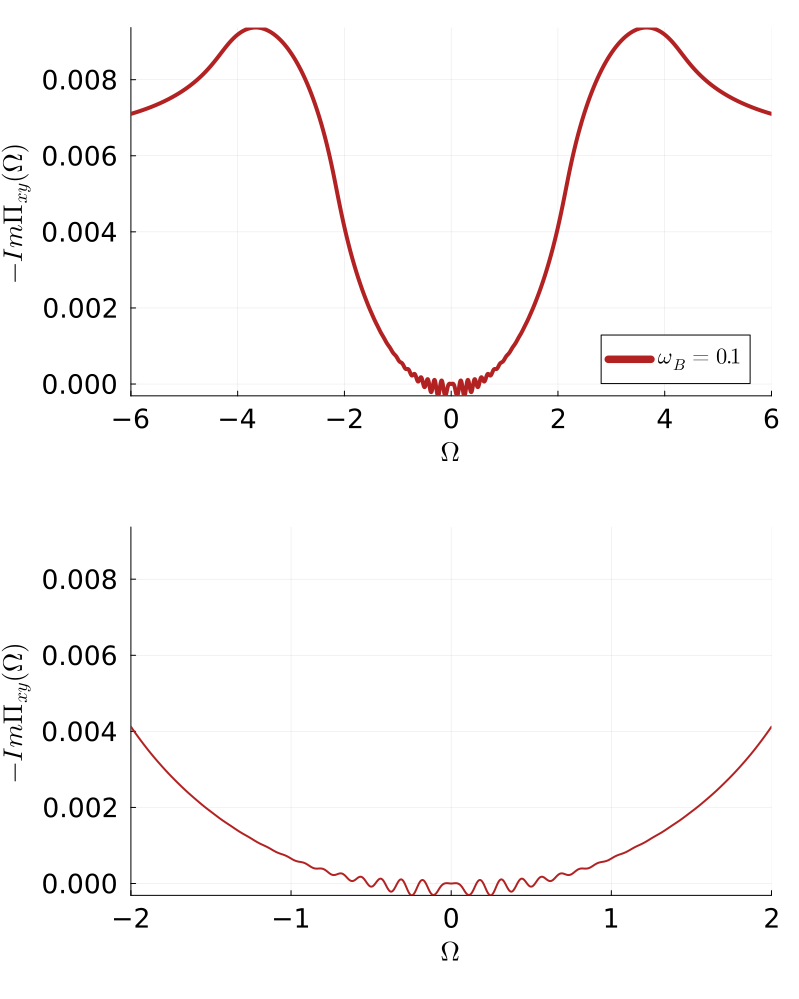}
	\caption{Numerical solution of $\Pi_{xy}$ when $\omega_B=0.1$. \label{fig:pixy}}
\end{figure}

The numerical solution obtained via the procedure mentioned above is illustrated by Fig.\ref{fig:gpropagator}-\ref{fig:pixy}. To make a straightforward comparison between our vector model and the scalar model, the parameters are set to be the same with those in ref.\cite{Guo2024} as follows: $W=4$, $\omega_B=0.1$, $k_F=1$, $\mu=0$, $m=1$, $\Lambda_q=2$, $K=1$ and $\omega\in[-16,16]$. In order that $n_t\leq n_+-1$, we require $W\geq 2\omega_B+k_F^2/m$. The result turns out to be qualitatively identical to that of the scalar case \cite{Guo2024}. 
Below we list our results. 

\begin{itemize}
	\item 
 Fig.\ref{fig:gpropagator} illustrates the behaviour of $-\omega_B\Im\bar{G}/\pi$, the electronic density of states \cite{Guo2024}
We compute $\bar{G}$ with various $\omega_B$ and Fig.\ref{fig:omegab} lists three examples where $\omega_B=0.1,0.05,0.02$. The solutions are featured by its oscillatory structure as a function of frequencies $\omega$. According to Fig.\ref{fig:gpropagator}, no oscillation is observed when $\omega_B=0$ (the grey line). 
Letting $\Delta\omega$ be the average separation between each peak for a given $\omega_B$, one finds $\Delta\omega$ is a linear function of $\omega_B$. The approximation that $\Delta\omega\simeq\omega_B$ becomes increasingly accurate as the magnetic field becomes stronger according to Fig.\ref{fig:level2}. Therefore these minima can be identified as Landau levels.   
	\item 
These oscillations can be understood as an analogue of de Haas-van Alphen effect \cite{Kittel1976,Ashcroft_Mermin_1988}, though the electrons in our model also interact with bosons. The frequency $\omega$ directly corresponds to the orbit in $k$-space and thus Landau levels. Suppose the area enclosed by the orbit is $\mathcal{S}$ in $k$-space, and for free electrons $\Delta \mathcal{S}=2\pi e B/\hbar$ gives the difference between successive orbitals. When $B$ is fixed, the peaks corresponds to the configurations where the intrinsic energy of the system takes an extreme value and the period is given by $\Delta \mathcal{S}$. Since we have bosons coupled to electrons, whose effect is encapsulated in $\Sigma$, the analysis for free electrons should be modified, and this can be seen from quantitative investigation. The oscillation amplitude is a result from competition between $|\Im{\Sigma}|$ and $\omega_B$ \cite{Guo2024}. According to the solution shown by Fig.\ref{fig:sigmaenergy}, within the bandwidth,  $|\Im{\Sigma}|$ increases with $|\omega|$. As a result in Fig.\ref{fig:omegab}, the oscillations become milder if $\omega_B$ is smaller or $|\omega|$ is larger.
To be more precise, let us take $\omega_B=0.1$ for instance. According to the solution in Fig.\ref{fig:sigmaenergy}, $|\Im{\Sigma(\omega)}|>\omega_B=0.1$ for $|\omega|\gtrsim 1$. Accordingly one finds in Fig.\ref{fig:gpropagator} that the oscillations are almost diminished when $|\omega|\gtrsim 1$ and the marginal-Fermi-liquid behaviour is more dominant in this region.  
	\item 
As for electron self-energy, our result also roughly matches well with zero-field solution ($\omega_B=0$) except small oscillations at low frequencies. Typically, within the bandwidth ($\omega\in[-2,2]$ approximately here), the self-energy goes linearly ($\Sigma\sim \omega\ln(\omega)$) like marginal fermi liquid when $\omega_B=0$ \cite{Ge:2024exw,Patel2022}. Moreover, since $\Sigma$ is a very small number, these oscillations for $\omega_B=0.1$ is suppressed if impurity scattering from potential disorder is introduced later, as potential disorder brings a much larger contribution to the total self-energy. Such an approximate linearity (inside the bandwidth) heralds the possible linear resistivity $\rho_{xx}$ in this model, as will be shown in the next section.
	\item 
The boson self-energy again roughly matches the zero-field solution \cite{Ge:2024exw,Patel2022}. This linearity is a consequence of fermion-boson spatially random coupling. It has been argued that at low temperature (or at low frequencies), such a bosonic self-energy makes the density of states $\sim T$ and the resistivity from bonson-electron scattering linear in $T$ \cite{Ge:2024exw}. Be along with what we have observed in electron self-energies $\Sigma$, it is reasonable to expect a linear resistivity at low temperature in our model. 
	\item 
We only show $\Pi_{xy}$ when $\omega_B=0.1$, because   $\Pi_{xy}$ does not exist when $ B=0$.  Its numerical value is much smaller than $\Pi_{xx}$, and this qualifies the approximations we made in eqn.\eqref{eqn:dxx} and eqn.\eqref{eqn:dxy}.
	\item 
The solutions go to zero outside the band ($|\omega|\gtrsim2$ for $W=4$), so only data of $\omega\in[-W/2,W/2]$ are reliable for further computation of transport properties.
\end{itemize}
\vskip 6cm 

\pagebreak


\pagebreak

\subsubsection{Potential disorder}
\begin{figure}[htbp]
	\centering
	\subfigure[Electron density of state.]{\includegraphics[width=0.45\textwidth]{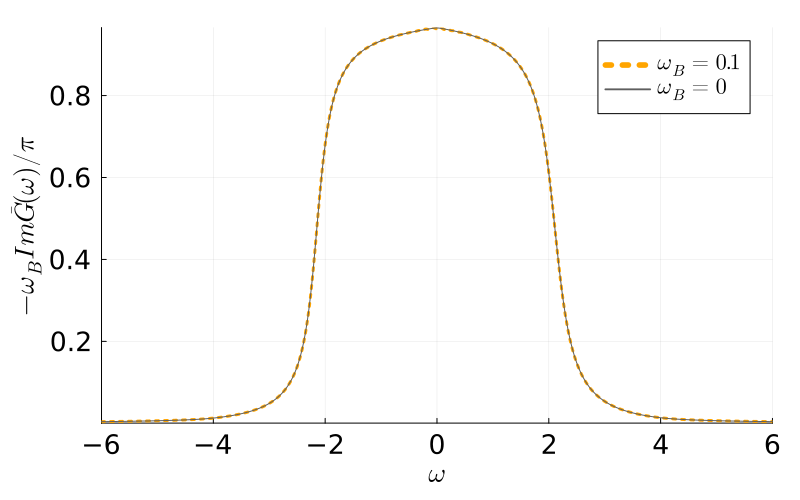}}
	\subfigure[Electron self-energy $\Sigma_K$.]{\includegraphics[width=0.45\textwidth]{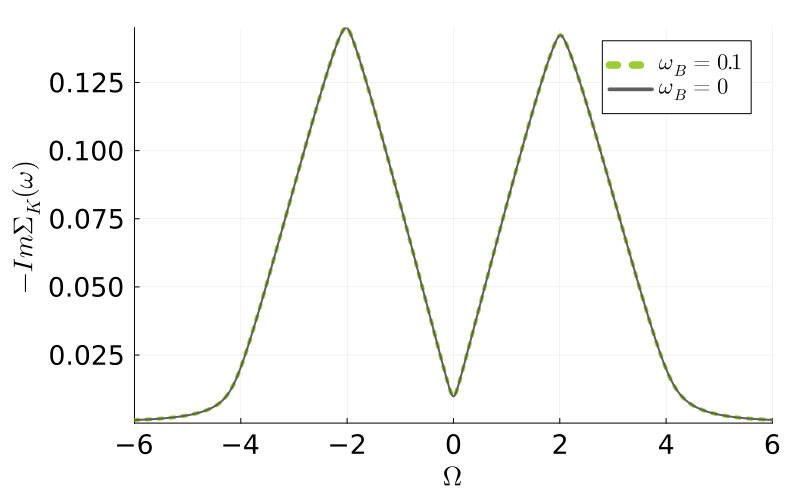}}	
	\subfigure[Boson self-energy $\Pi_{xx}$.]{\includegraphics[width=0.45\textwidth]{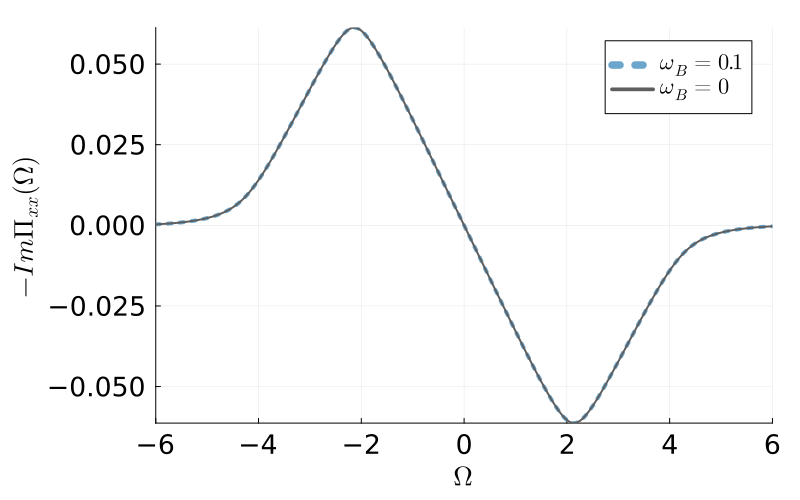}}	
	\subfigure[Boson self-energy $\Pi_{xy}$.]{\includegraphics[width=0.45\textwidth]{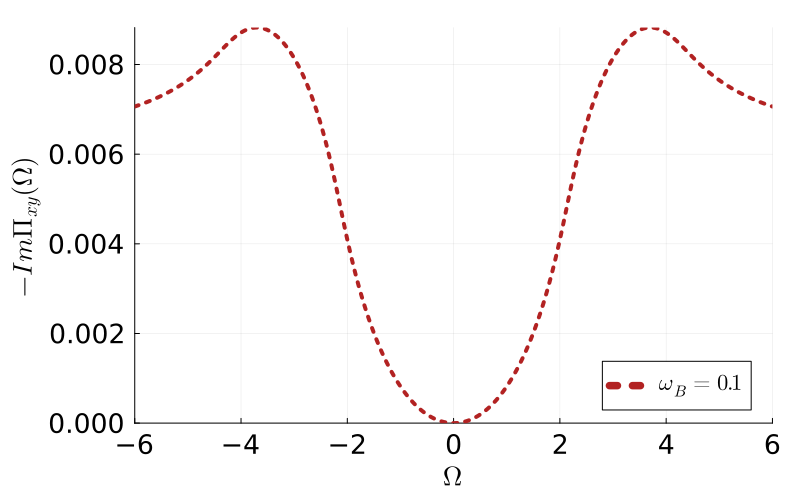}}	
	\caption{The solutions after potential disorder is introduced with $\Gamma=0.2$. The dotted lines represent solutions of $\omega_B=0.1$, and the solid lines shows the solutions of $\omega_B=0$. Two sets of solutions are almost the same, so we can assume potential disorder still contributes a constant self-energy. \label{fig:solution2}}
\end{figure}

In order to make comparisons with realistic materials, we now introduce a potential disorder $v_{ij}(\br)\psi_i^{\dagger}(\tau,\br)\psi_j(\tau,\br)$ which satisfies 
$$\langle v_{ij}(\br)\rangle=0, \quad \hbox{ and } \langle v_{ij}(\br) v^*_{i'j'}(\br')\rangle=v^2\delta(\br-\br')\delta_{ii',jj'}.$$ The total self-energy becomes
\begin{eqnarray}
	\Sigma(\omega)=\Sigma_v(\omega)+\Sigma_K(\omega),
\end{eqnarray} 
where $\Sigma_K$ is given by eqn.\eqref{eqn:f-selfenergy}, and 
\begin{eqnarray}\label{eqn:impurity}
	\Sigma_v(\bi\omega)&=&v^2\int\frac{d^3 \bk'}{(2\pi)^3}G(\bi\omega,\bk')\nonumber\\
	&=&v^2\sum_nG_n(\bi\omega)\int d\bx dk\psi_{nk}(\bx)\psi_{nk}^*(\bx)\nn
	&=&v^2g_L\sum_nG_n(\bi\omega).
\end{eqnarray}
This quantity is proportional to $\bar{G}$, and it becomes a constant when $\omega_B=0$, say 
\begin{eqnarray}
	\Sigma_v(\omega)=-\bi\frac{\Gamma}{2},
\end{eqnarray}
where $\Gamma\equiv v^2 m$ is the disorder scattering rate \cite{Ge:2024exw,Patel2022}. As is mentioned above, $\bar{G}$ has oscillations and the oscillations exponentially vanish once $\Im{\Sigma(\omega)}>\omega_B$. In this sense, we can approximately take
\begin{eqnarray}\label{eqn:appself}
	\Sigma(\omega)\simeq -\bi\frac{\Gamma}{2}+\Sigma_K(\omega),
\end{eqnarray}
by choosing $\Gamma>\omega_B$. With the same parameters chosen in section \ref{subsec:solution} and taking $\Gamma=0.2$, Fig.\ref{fig:solution2} verifies the self-consistency of assumption \eqref{eqn:appself}. That $\Sigma_v$ is a constant means $\bar{G}|_{\omega_B\neq 0}$ needs to take a similar form with $\bar{G}|_{\omega_B= 0}$, since the oscillatory structure is diminished by potential disorder. Fig.\ref{fig:solution2} precisely shows the similarity between the solutions of $\omega_B=1$ and  $\omega_B=0$. Again, with a cutoff from bandwidth $W$, one finds only when $\omega\in[-W/2,W/2]$ can Eqn.\eqref{eqn:impurity} be approximately a constant, and $\Sigma_v$ drops to zero when the frequency goes beyond the bandwidth. 
\\ 

\section{Transport}
In this  section we numerically compute the transport properties of random vector model (with potential disorder) in a magnetic filed. We will obtain conductivity, Hall angle, and resistivity step by step. Letting $\mathscr{A}$ be the quantum part of the external field, one can compute the conductivity $\sigma^{\mu\nu}$ from its polarisation bubbles $\tilde{\Pi}^{\mu\nu}$ via \emph{Kubo formula} in real frequencies
\begin{eqnarray}\label{eqn:kubo}
	\sigma^{\mu\nu}(\Omega)=-\frac{e^2}{S}\frac{\tilde{\Pi}^{\mu\nu}(\Omega+\bi0^+)-\tilde{\Pi}^{\mu\nu}(0)}{\bi\Omega}.
\end{eqnarray}
The \emph{Hall angle} is defined as \cite{Schrieffer2007}
\begin{eqnarray}\label{eqn:hallangle}
	\tan(\Theta_H)\equiv\frac{\sigma_{xy}}{\sigma_{xx}}.
\end{eqnarray}
The resistivity will also be computed from the conductivity, and we will focus on $\rho_{xx}$,
\begin{eqnarray}\label{eqn:resistivity}
	\rho_{xx}=\frac{\sigma_{xx}}{(\sigma_{xx})^2+(\sigma_{xy})^2}.
\end{eqnarray}\\

\begin{figure}[htbp]
	\centering
	\includegraphics[width=0.25\textwidth]{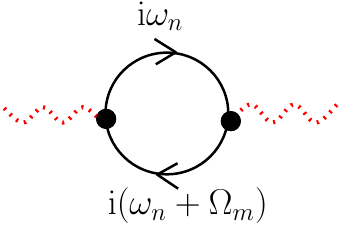}
		\caption{Bare current-current correlation. It yield a Drude-like contribution to the conductivity $\sim 1/(\bi\Omega+\Gamma)$.\label{fig:polar_diagram}}
\end{figure}

\begin{figure}[htbp]
	\centering
	\subfigure[Self-energy contribution]{\includegraphics[width=0.3\textwidth]{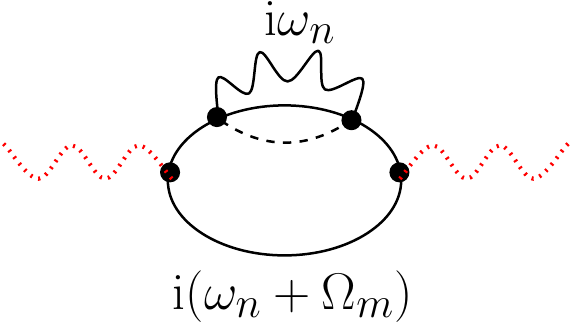}}
    \subfigure[MT diagram]{\includegraphics[width=0.3\textwidth]{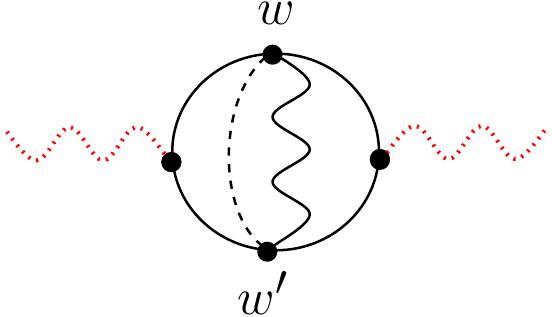}}	
	\caption{The MT diagram will precisely cancel the contribution from electron self-energy $\Sigma_K$. \label{fig:MT}}
\end{figure}

\begin{figure}[htbp]
	\centering
	\includegraphics[width=0.4\textwidth]{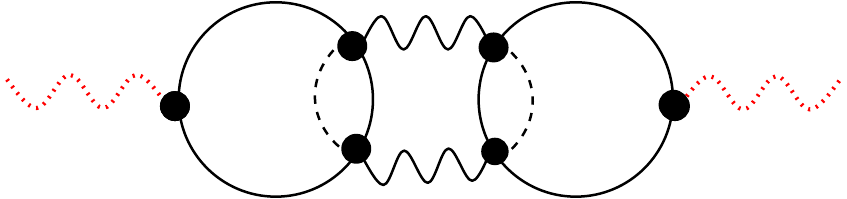}
	\qquad
	\includegraphics[width=0.4\textwidth]{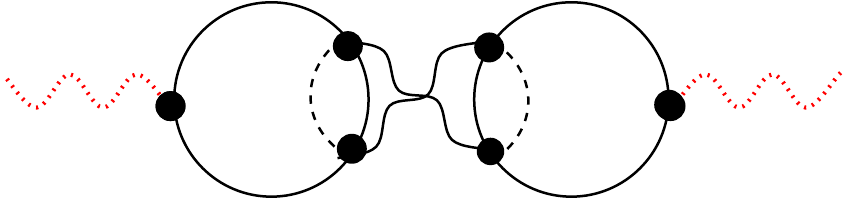}
	\caption{AL diagrams vanish due to spatial delta.\label{fig:AL_diagram}}
\end{figure}

\begin{figure}[htbp]
	\centering
	\includegraphics[width=.3\textwidth]{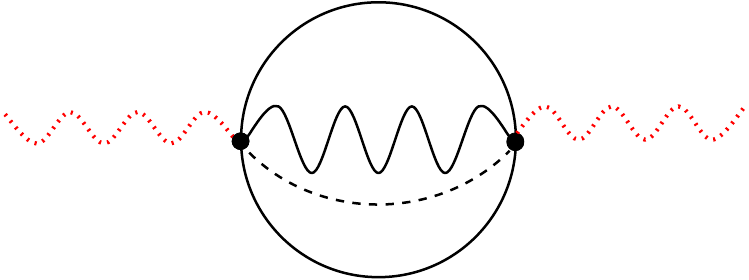}
	\caption{The polarisation bubble contributing to the conductivity. The wavy line represents vector-field propagator and the dashed line means spatial contraction between two vertices. \label{fig:polar}}
\end{figure}

Knowing $\sigma_{\mu\nu}$ enables us to know other two quantities, so we can first compute polarisation bubbles of $\mathscr{A}$ and use the Kubo formula to find the conductivity. We will compute the polarisation as an expansion of $K^2$ up to the order $\mathcal{O}(K^2)$.
In principle, there are three types of Feynman diagrams to be computed. 
First, a bare polarisation bubble, the simplest diagram, given by Fig. \ref{fig:polar_diagram}, will yield a residual conductivity $\sigma_0(\Omega)$ originating from potential disorder. The dotted wavy red lines stand for propagators of the external field, $\langle\mathscr{A}^{\mu}(\tau,\bx)\mathscr{A}^{\nu}(\tau',\bx')\rangle$.  Since the off-diagonal term is too small compared with diagonal term and our numerical solution is almost the same as zero-field solutions because of potential disorder, we can approximately take
\begin{eqnarray}
	\sigma_0(\Omega)\simeq\frac{e^2k_F^2\mathcal{N}}{2m^2}\frac{1}{\Gamma+\bi\Omega}
	\begin{pmatrix}
		1 & 0\\
		0 & 1
	\end{pmatrix}.
\end{eqnarray}
	This term becomes a constant after we take its real part, so the non-trivial $\Omega$-dependence comes from higher-order graphs and we will consider the Feynman diagrams up to two loops. 
	The bare polarisation receives corrections from boson-fermion self-energy $\Sigma_K(\Omega)$ and vertex corrections named Maki-Thompson (MT) diagrams, as are shown in Fig.\ref{fig:MT}. For the scalar coupling \cite{Patel2022}, MT diagram is vanishing, so $\Sigma_K(\Omega)$ accounts for the linear-$T$ resistivity. In our vector model, however, MT graph is non-zero and it will cancel the contribution from $\Sigma_K(\Omega)$. In addition, thanks to spatial delta introduced from spatial randomness $K_{ijl}(\br)$, vertex correction named Aslamazov-Larkin graphs, shown in Fig.\ref{fig:AL_diagram}, vanish. 
Consequently, the non-trivial conductivity will be contributed by a bubble illustrated in Fig.\ref{fig:polar}, and this bubble is from the coupling $\mathscr{A}^{\mu}\ba_{\mu}\psi^{\dagger}\psi$. This polarisation reads
\begin{eqnarray}\label{eqn:polarisation2}
	&&\tilde{\Pi}^{\nu\mu}(\bi\Omega)
	=-\frac{S}{2m^2}T^2\sum_{\omega,\Omega'}\sum_{n,n'}\int \frac{d^2\bq}{(2\pi)^2}G_{n}(\bi\omega)
	G_{n'}(\bi(\Omega+\omega+\Omega'))D^{\mu\nu}(\bi\Omega',\bq).
\end{eqnarray}

This bubble can be computed numerically using the the solutions to Dyson's equation, whose process has been illustrated in section \ref{subsec:solution}. 
Defining
\begin{eqnarray}
	&&\tilde{G}(\omega)=-2n_F(\omega)\Im\bar{G}(\omega),\\
	&&\tilde{D}_{\mu\nu}(\Omega)=-2n_B(\Omega)\Im\bar{D}_{\mu\nu}(\Omega),\\
	&&\mathcal{G}(\Omega)=-2n_F(\omega)\bar{G}(\omega),
\end{eqnarray}
we perform analytical continuation $\bi\Omega\to\Omega+\bi\eta$ and find \eqref{eqn:polarisation2} becomes
\begin{eqnarray}\label{eqn:polart}
	\tilde{\Pi}_{\nu\mu}(t)
	&=&\frac{S}{2m^2}\Big(\bar{G}(t)\tilde{G}(-t)\tilde{D}_{\mu\nu}(-t)
	+\tilde{G}(t)\tilde{G}(-t)\bar{D}_{\mu\nu}(-t)\nn
	&&-\mathcal{G}(-t)\bar{G}(t)\tilde{D}_{\mu\nu}(-t)
	+\mathcal{G}(-t)\tilde{G}(t)\Im\bar{D}_{\mu\nu}(-t)\Big),
\end{eqnarray}
in real time representation \footnote{More details can be found in Appendix \ref{app:summation}.}. Substituting $G(t)$ and $D_{\mu\nu}(t)$ into Eqn.\eqref{eqn:polart} and doing Fourier transformation, we can obtain $\tilde{\Pi}_{\mu\nu}(\Omega)$. Applying the Kubo formula \eqref{eqn:kubo}, one obtains the conductivity $\sigma_{\mu\nu}$.\\

\begin{figure}[htbp]
	\centering
	\subfigure[The longitudinal conductivity $\sigma_{xx}(\Omega)$]{\includegraphics[width=0.8\textwidth]{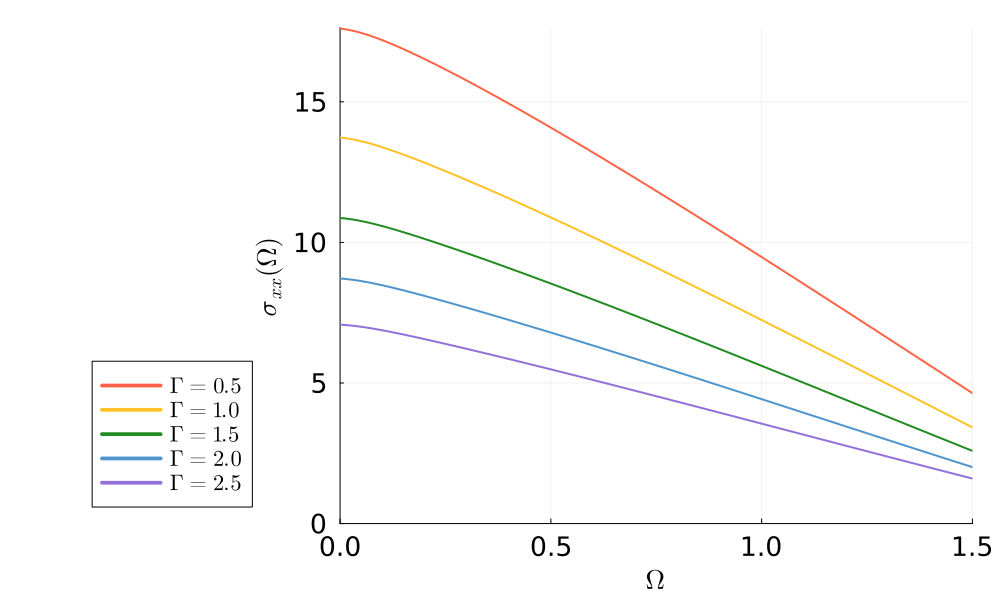}}
	\subfigure[The Hall conductivity $\sigma_{xy}(\Omega)$]{\includegraphics[width=0.8\textwidth]{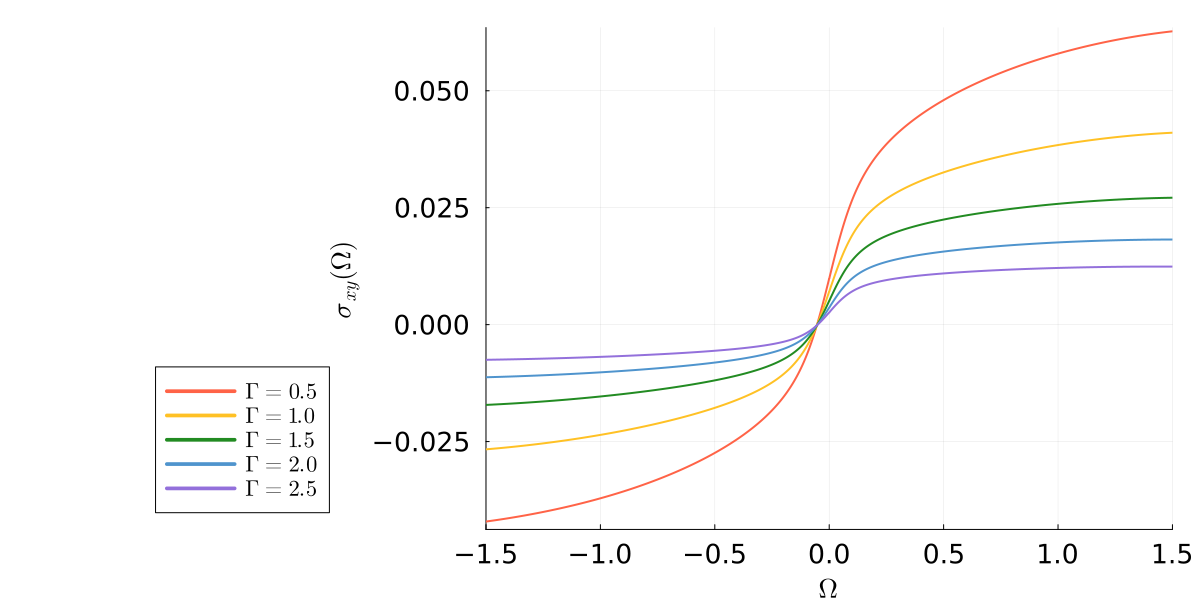}}	
	\caption{Conductivity $\sigma_{xx}(\Omega)$ and $\sigma_{xy}(\Omega)$ with various impurity scattering rate $\Gamma$. The red, yellow, green, blue, and purple lines represents $\Gamma=0.5, 1.0, 1.5, 2.0, 2.5$ respectively. The conductivity declines as $\Gamma$ increases. \label{fig:conductivity}}
\end{figure}

Fig.\ref{fig:conductivity} shows the longitudinal conductivity $\sigma_{xx}(\Omega)$ and Hall conductivity $\sigma_{xx}(\Omega)$ at zero temperature. The parameters are the same with those in \ref{subsec:solution}. The dependence on $\Omega$ can be translated to the dependence on $T$ for DC conductivities \cite{Patel2022}. Conductivities with various $\Gamma$ are investigated, and one finds conductivity is larger when $\Gamma$ is smaller. Moreover, the conductivities of the vector model in this article is qualitatively the same with the scalar model studied in \cite{Guo2024}, so spatially random couplings between fermions and bosons may generally share the transport properties regardless of the coupling type.\\

\begin{figure}[htbp]
	\centering
	\includegraphics[width=.6\textwidth]{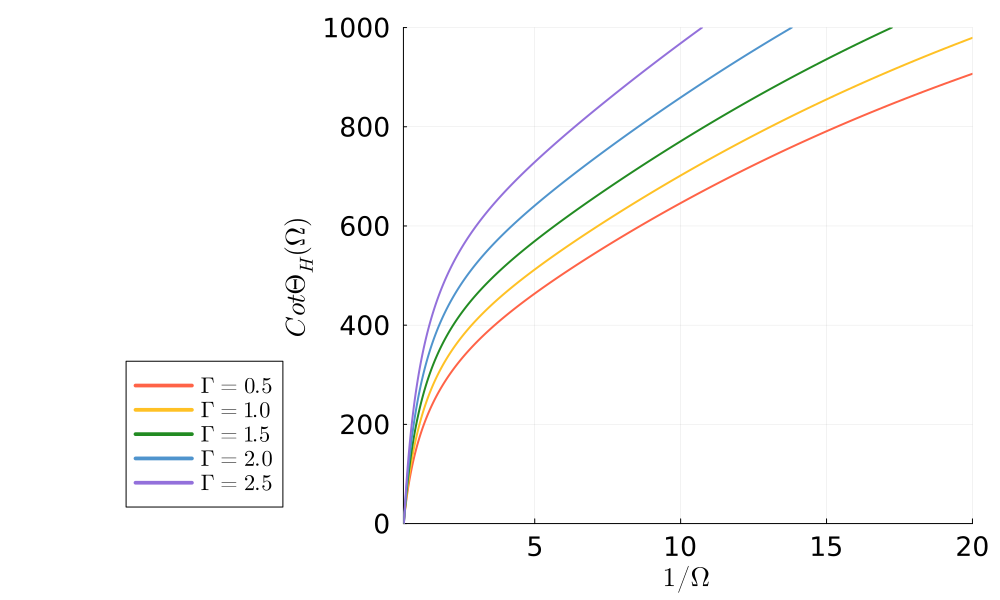}
	\caption{Hall angle $\cot(\Theta_H)$ obtained from the data in Fig.\ref{fig:conductivity}, and one finds $\coth(\Theta_H)$ is approximately a linear function of $1/\Omega$. This implies that  $1/T$ behaviour is found  instead of $T^2$. The red, yellow, green, blue, and purple lines represents $\Gamma=0.5, 1.0, 1.5, 2.0, 2.5$ respectively, and the slope is almost independent of $\Gamma$. \label{fig:hallangle}}
\end{figure}

Having found the conductivity, one can directly move on to Hall angle defined by Eqn.\eqref{eqn:hallangle}. Substituting the data in Fig.\ref{fig:conductivity} into $\cot(\Theta_H)=\sigma_{xx}/\sigma_{xy}$, one finds that $\cot(\Theta_H)$ has no $T^2$-dependence. In fact, the behaviour cannot match none of the strange metals to the best of our knowledge. According to Fig.\ref{fig:conductivity}, one finds in our model, the scattering rate $\tau\sim T$. In usual systems, one expects $\sigma_{xx}/\sigma_{xy}=\rho_{xx}/\rho_{xy}\sim \tau^{-1}\sim 1/T$, and Fig.\ref{fig:hallangle} shows a $1/T$ scaling behaviour. There are thus no anomalies in the Hall angle in this article. Indeed, the Hall angle is not `strange' at all since it strictly obeys the analysis for a normal system.
Therefore, it seems that action \eqref{eqn:action} cannot account for the Hall angle of any strange metal material observed so far \cite{Schrieffer2007}.\\

\begin{figure}[htbp]
	\centering
	\includegraphics[width=.6\textwidth]{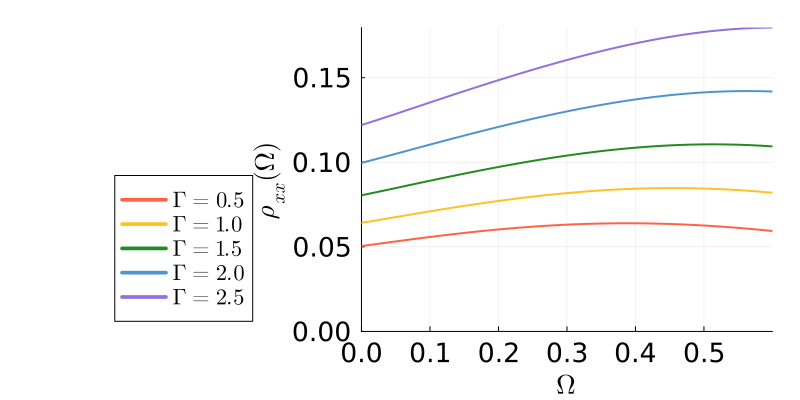}
	\caption{Resistivity $\rho_{xx}$using data from Fig.\ref{fig:conductivity}. The linear resistivity (at low temperatures) is preserved in the presence of a magnetic field. \label{fig:resistivity}}
\end{figure}

Finally using eqn.\eqref{eqn:resistivity}, we can get the resistivity $\rho_{xx}$, as is shown by Fig. \ref{fig:resistivity}. Linear-$T$ conductivity yields linear-$T$ resistivity at low temperatures, and the linear resistivity also exists in the model without a magnetic field \cite{Ge:2024exw,Patel2022}. Additionally, as temperature goes higher, the linearity will disappear. This is because the conductivity $\sigma_{xx}\simeq A-BT$ and
\begin{eqnarray}\label{eqn:app}
	\rho_{xx}\simeq\frac{1}{\sigma_{xx}}=\frac{1}{A-BT}\simeq \frac{1}{A}+\frac{B}{A^2}T
\end{eqnarray} 
only when $|BT|\ll |A|$ \cite{Ge:2024exw,Patel2022}. Therefore, the linearity disappears when temperature is too high. Suppose the resistivity is linear in $T$ for $T<T_L$.
Though the numerical value cannot be specified from Fig. \ref{fig:resistivity}, we find that the larger $\Gamma$ is, the higher is $T_L$. This property is what we can predict from Eqn.\eqref{eqn:app}. Admittedly, this reflects a crucial limitation of the SYK-rised models, either scalar \cite{Patel2022} or vector coupling, as observations have shown several examples wherein $A<BT$ \cite{PhysRevLett.59.1337,Hussey2013,Cooper2009,PhysRevB.95.224517}. An important task for the future is to identify a mechanism that holds at higher temperatures as well.\\

\section{Discussion}

Previous researches have shown that spatially random coupling between a fermi surface and boson field yields linear-$T$ resistivity at low temperatures (in $(2+1)$ dimensions) \cite{Ge:2024exw, Patel2022}. In order to check whether the spatial randomness can reproduce other properties of a strange metal and find possible realistic examples it describes, one adds an external magnetic field and studies the transport properties in a vector model introduced in \cite{Ge:2024exw}. 
After numerically solving Dyson's equations, we find the longitudinal conductivity $\sigma_{xx}$ and Hall conductivity $\sigma_{xy}$. The Hall angle turns out to have no correspondence with experimental observations, but linear $\rho_{xx}$ remains in this system.\\

The `non-strange' Hall angle in this system implies that the spatially random coupling may not be the elixir of strange metal, but the preserved linear-$T$ resistivity in a magnetic field further verifies that such a spatial randomness can be the mechanism for least the linearity at low temperatures. According to the analysis in \cite{Ge:2024exw}, the longitudinal resistivity results from boson-electron scatterings. Because the Dyson's equations keep qualitatively the same solutions, the same argument applies in this article again. Here we briefly summerise the analysis, and the full discussion can be found in \cite{Ge:2024exw}. As we find in Fig.\ref{fig:solution2}, the boson self-energy $\Pi_{xx}\sim \Omega$ if potential disorder is included in our system. As a result, $\Omega\sim \bq^2$ at low frequencies, and the bosonic density of states reads
\begin{equation}\label{eqn:dos}
	\int d^d\bq\frac{1}{e^{\beta\Omega}-1}\sim T^{d/2}\int dx\frac{x^{(d-2)/2}}{e^x-1},
\end{equation}
by taking $\beta\Omega =x$. Eqn.\eqref{eqn:dos} shows the scaling behaviour of the resistivity caused by boson-electron scatterings. Meanwhile, the spatial randomness relaxes the momentum conservation on electron-boson interaction vertices, so there will be no small-angle correction $(1-\cos(\theta))\sim T$, with $\theta$ the scattering angle. Consequently according to eqn.\eqref{eqn:dos}, without small-angle correction, the overall resistivity $\rho_{xx}$ will be linear in $T$ when $d=2$. The linear-$T$ appears again in the presence of a magnetic field, indicating that the success of spatially random coupling in previous research \cite{Patel2022, Ge:2024exw} may not simply be a fluke.\\

As for Hall conductivity, however, the mechanism may be more complex, especially when this anomaly is not as universal as linear resistivity (though the quadratic behaviour can be reproduced from the holographic method \cite{Ge:2016lyn,Ge:2016sel,Ge:2019fnj}). A promising resolution is to include spins. For example, Anderson suggests that the spinon-spinon interaction could be responsible for the scaling behaviour of $\sigma_{xy}$ \cite{Anderson1991}. Furthermore, the Curie-Weiss law requires a magnetic susceptibility $\chi\sim 1/T^{\gamma}$, with $\gamma$ the critical exponent. In \cite{Kontani2007}, the magnetic susceptibility will modify the Hall conductivity such that $\sigma_{xy}\sim \chi \cdot \tau^2$, making it possible to obtain $\cot(\Theta_H)\sim T^2$. 
In contrast, only spinless particles are considered so far in both scalar model \cite{Patel2022} and vector model \cite{Ge:2024exw}. Therefore, more interaction types may need to be considered to build a full theory of strange metal on a firm ground.

\appendix
\section{Matsubara Summation}\label{app:summation}
This section illustrates how to apply the standard contour integral technique to compute a Matsubara summation \cite{Abrikosov1965}, whose result is used in numerical computation in section \ref{subsec:solution}. \\

\begin{figure}[htbp]
	\centering
	\includegraphics[width=.65\textwidth]{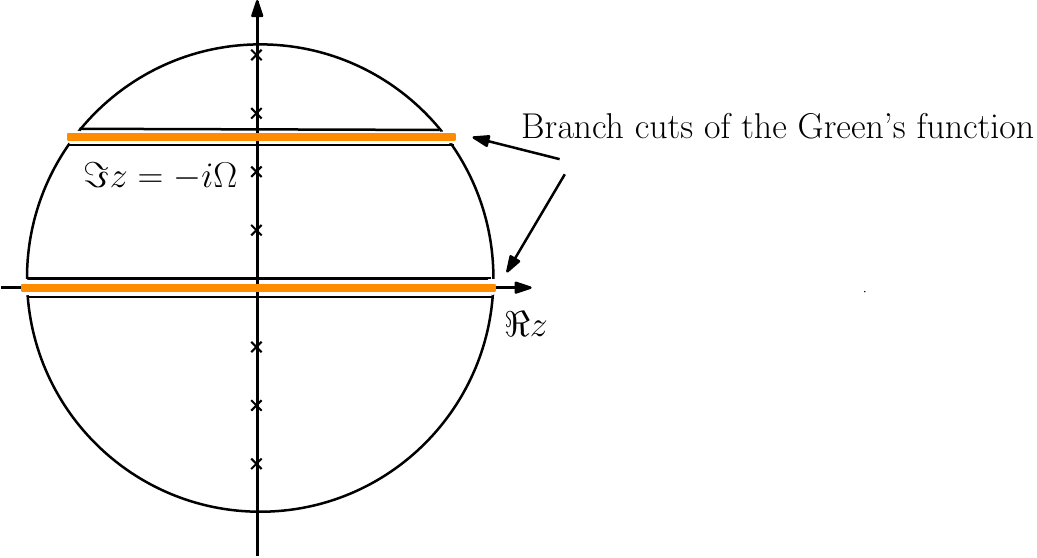}
	\caption{Branch cuts and poles.\label{fig:contour1}}
\end{figure}

A common way to do a fermionic Matsubara summation of fermions is to evaluate a integral \cite{Altland2023}
\begin{eqnarray}
	I=\lim_{R\to\infty}\oint \frac{dz}{2\pi\bi}n_F(z)f(z),
\end{eqnarray}
where 
\begin{eqnarray}
	f(z)= \xi(z)\chi(z+\bi\Omega).
\end{eqnarray}
The function $f(z)$ has branch cuts along $\Im{z}=0$ and $\Im{z}=-\Omega$. Here $\Omega=2n\pi T$ is a Boson Matsubara frequency. Functions $\xi$ and $\chi$ are Green's functions in this paper. The fermi function $n_F$ brings poles at $z=\bi (2m+1)\pi T$ with $n\in\mathbb{Z}$.
We need to deform the contour integral in order to avoid the cuts \cite{Abrikosov1965,Schmalian1996}. As is illustrated in Fig.\ref{fig:contour1}, the contour is divided by four horizontal lines.

The integrand vanishes at infinity, so only integrals along four horizontal lines contribute. Applying residue theorem yields
\begin{eqnarray}\label{eqn:summation1}
	&&-T\sum_m f(\bi\omega_m)\nn
	&=&\int_{-\infty}^{+\infty}\frac{dz}{2\pi\bi}\Big(
	n_F(z)\xi(z+\bi\eta')\chi(z+\bi\Omega)-n_F(z)\xi(z-\bi\eta')\chi(z+\bi\Omega)\nn
	&&+n_F(z)\xi(z-\bi\Omega)\chi(z+\bi\eta')-n_F(z)\xi(z-\bi\Omega)\chi(z-\bi\eta')
	\Big)\nn
	&=&2\int_{-\infty}^{+\infty}\frac{dz}{2\pi}\Big(n_F(z)\Im{\xi(z+\bi\eta')}\chi(z+\bi\Omega)
	+n_F(z)\xi(z-\bi\Omega)\Im{\chi(z+\bi\eta')}\Big).
\end{eqnarray}
Here we use the fact that both $\chi$ and $\xi$ satisfy $\chi_R(\omega)/\xi_R(\omega)=\chi_A^*(\omega)/\xi_A^*(\omega)$ after an analytical continuation $\bi\omega\to\omega\pm\bi 0^+$, where $R$ and $A$ refer to `retarded' and `advanced' respectively.\\

Using the Fourier transform $F(t)=\int d\omega f(\omega)\exp(-\bi\omega t)/2\pi$ and the convolution theorem, one finds the result in real-time representation,
\begin{eqnarray}\label{eqn:fermionsum}
	F(t)=-\tilde{\xi}^*(t)\chi(t)+\tilde{\chi}(t)\xi(-t),
\end{eqnarray}
where
\begin{align}
	&\tilde{\xi}(\omega)=-2n_F(\omega)\Im{\xi(\omega)},\\
	&\tilde{\chi}(\omega)=-2n_F(\omega)\Im{\chi(\omega)}.
\end{align}\\

The summation of bosonic Matsubara frequency can be evaluated in the same way by computing
\begin{eqnarray}
	I=\lim_{R\to\infty}\oint \frac{dz}{2\pi\bi}n_B(z)g(z),
\end{eqnarray}
where the function 
\begin{eqnarray}
	g(z)\equiv \Xi(z)\chi(z+\bi\omega)
\end{eqnarray}
also has branch cuts along $\Im{z}=0$ and $\Im{z}=-\Omega$, with $\omega=2n\pi T$ a fermion frequency.
The same contour integral in Fig.\ref{fig:contour1} can be used to avoid the cuts. Now one finds poles at $z=\bi(2m)\pi T$, and
\begin{eqnarray}\label{eqn:summation2}
	&&T\sum_mf(\bi\Omega_m)\nn
	&=&\int_{-\infty}^{\infty}\frac{dz}{2\pi \bi}
	\Big(
	n_B(z)\Xi(z+\bi\eta')\chi(z+\bi\omega)-n_B(z)\Xi(z-\bi\eta')\chi(z+\bi\omega)\nn
	&&+n_B(z-\bi\omega)\Xi(z-\bi\omega)\chi(z+\bi\eta')-n_B(z-\bi\omega)\Xi(z-\bi\omega)\chi(z-\bi\eta')
	\Big)\nn
	&&-n_F(z)\xi(z-\bi\omega)\chi(z+\bi\eta')+n_F(z)\xi(z-\bi\omega)\chi(z-\bi\eta')
	\Big)\nn
	&=&2\int_{-\infty}^{\infty}\frac{dz}{2\pi}\Big(n_B(z)\Im{\Xi(z+\bi\eta')}\chi(z+\bi\omega)-
	n_F(z)\Im{\chi(z+\bi\eta')}\Xi(z-\bi\omega)\Big).
\end{eqnarray}
This convolution structure gives the Fourier transform of $g(z)$,
\begin{eqnarray}\label{eqn:bosonsum}
	G(t)=-\tilde{\Xi}^*(t)\chi(t)-\tilde{\chi}(t)\Xi(-t),
\end{eqnarray}
where
\begin{align}
	&\tilde{\Xi}(\omega)=-2n_B(\omega)\Im{\Xi(\omega)},\\
	&\tilde{\chi}(\omega)=-2n_F(\omega)\Im{\chi(\omega)}.
\end{align}
In many cases, it is formidable to evaluate equations such as \eqref{eqn:summation1} and \eqref{eqn:summation2} directly. It is more convenient to work in time representation first and move back to frequency representation via fast Fourier transform. Therefore, Eqn.\eqref{eqn:fermionsum} and eqn.\eqref{eqn:bosonsum} will be useful when one tries finding numerical solutions. 

\section{data}
All the data used in this article can be found in  \cite{data}. 

\acknowledgments
Authors would like to thank Yu-Ge Chen for inspiring discussion on numerical calculation.
This work is partly supported by NSFC, China (Grant No. 12275166 and No. 12311540141). This work is also supported by NRF of Korea with grant No. NRF-2021R1A2B5B02002603, RS-2023-00218998 and NRF-2022H1D3A3A01077468.


\bibliographystyle{JHEP}
\bibliography{library.bib}

\end{document}